\documentclass[preprint,10pt,notitlepage,oneside,aps,twocolumn]{revtex4}%
\usepackage{amsfonts}
\usepackage{amsmath}
\usepackage{amssymb}
\usepackage{graphicx}%
\begin{document}
\title{A semiclassical derivation of Einstein's rate equations for light--matter interaction}
\author{Germ\'{a}n J. de Valc\'{a}rcel and Eugenio Rold\'{a}n}
\affiliation{Departament de \'{O}ptica, Universitat de Val\`{e}ncia, Dr. Moliner 50, 46100
Burjassot, Spain}

\begin{abstract}
Einstein's rate equations are derived from the semiclassical Bloch equations
describing the interaction of a classical broadband light field with a
two--level system.

\end{abstract}
\maketitle

\section{Introduction}

Lorentz's damped and forced harmonic oscillator equation \cite{Lorentz} and
Einstein's rate equations \cite{Einstein} are very valuable and widely used
heuristic models for the description of light--matter interaction. Although
each of these models applies to different situations (roughly speaking, the
Lorentz equation usually applies when weak or highly detuned fields are
involved whilst Einstein rate equations are most adequate for describing
resonant interaction of atoms with broadband fields), both can be fully
justified withing the framework of quantum optics in the appropriate limits
\cite{Milonni,Shore,Loudon,Dodd}.

Generically speaking, formal derivations from first principles of heuristic
models are important. This is not only for easthetical reasons or completitude
arguments as these derivations allow to clarify the domain of applicability of
the heuristic model. The situation of the Lorentz and Einstein models is quite
different in this sense: Whilst formal derivations of the Lorentz equation
from the semiclassical theory of light--matter interaction, \textit{i.e.} from
the optical Bloch equations, can be found in several textbooks (see, e.g.
\cite{Milonni,Shore}), formal derivation of the Einstein's rate equations (ERE
for short) cannot be found in textbooks nor in research articles to the best
of the authors knowledge \cite{note0}. This is the most surprinsing given the
paramount importance of the Einstein model.

Of course, the above does not mean at all that the connection between the
optical Bloch equations and ERE has not been treated: Several quantum optics
textbooks (see, e.g. \cite{Shore,Loudon,Dodd}) discuss ERE and provide
derivations of the $A$ and $B$ Einstein coefficients. We find it important to
remark that whilst the derivation of the $A$ coefficient requires quantization
of both matter and electromagnetic field (see, e.g. \cite{Loudon})
\cite{note1}, the $B$ coefficient can be derived within the standard
semiclassical formalism (in which only matter is quantized), a derivation that
provides the same result as that obtained with the fully quantized theory
\cite{Loudon}. Obviously, the values of $A$ and $B$ thus obtained verify the
relation already given by Einstein for the ratio $A/B$. It is also worth
noting that the semiclassical derivations of the $B$ coefficient make use of
the assumption of weak field, an approximation that is not necessary indeed.

In this article we derive ERE from the Bloch equations with heuristically
added spontaneous emission terms as we are interested in stimulated processes.
The article is organized as follows: In section II we present briefly ERE; in
section III we introduce the optical Bloch equations and reduce them to an
integro-differential equation for the atomic inversion. Then, we derive ERE,
first in the limit of weak field intensity (section IV) and later removing
this assumption (section V). We find it convenient to make this distinction
because the derivation in the weak field limit does not involve certain
statistical assumptions which are necessary for arbitrary field strength. In
section VI we discuss the limits of validity of ERE and finally, in section
VI, we give our main conclusions.

\section{Einstein rate equations}

ERE were postulated by Albert Einstein in his famous 1917 paper
\cite{Einstein} and a clear presentation of them can be found, e.g., in
\cite{Loudon}. Einstein introduced three basic processes describing the
interaction of a broadband isotropic light field with an atomic two--level
system: Stimulated absortion, stimulated emission, and spontaneous emission.
Einstein assigned to these processes a constant probability per unit time
given by $B_{12}W\left(  \omega_{21}\right)  ,B_{21}W\left(  \omega
_{21}\right)  ,$ and $A_{21}$, respectively, being $W\left(  \omega
_{21}\right)  $ the spectral energy density of the light field at the
frequency $\omega_{21}$ of the atomic transition, and $A_{21},B_{12},$ and
$B_{21}$ the Einstein coefficients, which do not depend on the field strength
nor on time.

Assuming that all $N$ atoms are in either the upper excited state (labelled
$2$) or in the lower fundamental state ($1$), the time evolution of the
population of atoms in level $2$ ($N_{2}$) and $1$ ($N_{1}=N-N_{2}$) are
governed by
\begin{align}
\dot{N}_{2}  &  =-A_{21}N_{2}+W\left(  \omega_{21}\right)  \left(  B_{12}%
N_{1}-B_{21}N_{2}\right)  ,\label{Einstein1}\\
\dot{N}_{1}  &  =-\dot{N}_{2}, \label{Einstein2}%
\end{align}
assuming that the number of atoms $N$ is large enough for the individual
absorptions and emissions produce smooth temporal changes in the populations
(the dot stands for time derivative).

Through the analysis of thermal equilibrium it is easy to prove that
$B_{12}=B_{21}$ and $A_{21}/B_{21}=\hbar\omega_{21}^{3}/\pi^{2}c^{3}$
\cite{Loudon}. Finally, by introducing the normalized population inversion
\begin{equation}
\bar{n}=\frac{N_{2}-N_{1}}{N},
\end{equation}
($1\geq\bar{n}\geq-1$) we can rewrite ERE Eqs.(\ref{Einstein1}) and
(\ref{Einstein2}) in the simpler form
\begin{equation}
\frac{d}{dt}\bar{n}\left(  t\right)  =-A\left[  \bar{n}\left(  t\right)
+1\right]  -2BW\left(  \omega_{21}\right)  \bar{n}\left(  t\right)  ,
\label{Einstein}%
\end{equation}
with $A\equiv A_{21}$ and $B\equiv B_{12}=B_{21}$.

\subsection{\label{weak}Weak field limit of ERE}

For later purpouses it is convenient to consider the special case of a weak
light field. This can be done by writing $W\left(  \omega_{21}\right)
=\varepsilon W_{1}\left(  \omega_{21}\right)  $, with $\varepsilon\ll1$ and
$W_{1}\left(  \omega_{21}\right)  $ an order one quantity, expanding the
inversion in powers of $\varepsilon$
\begin{equation}
\bar{n}\left(  t\right)  =\sum_{m=0}^{\infty}\varepsilon^{m}\bar{n}_{m}\left(
t\right)  , \label{desarrollo}%
\end{equation}
and writing the equation of evolution for the first orders in $\varepsilon$.
At order $\varepsilon^{0}$ Eq.(\ref{Einstein}) reads $d\bar{n}_{0}%
/dt=-A\left(  \bar{n}_{0}+1\right)  $, and thus $\bar{n}_{0}=-1$ after a
transient. Making use of this, at order $\varepsilon$ one gets
\begin{equation}
\frac{d}{dt}\bar{n}_{1}\left(  t\right)  =-A\left[  \bar{n}_{1}\left(
t\right)  +1\right]  +2BW_{1}\left(  \omega_{21}\right)  ,
\label{Einstein weak}%
\end{equation}
which is the weak field limit of ERE.

\section{Bloch equations for a two--level system}

Let us consider a light field of the form
\begin{align}
\mathcal{\vec{E}}\left(  \mathbf{r},t\right)   &  =\mathbf{E}\left(
\mathbf{r},t\right)  +\mathbf{E}^{\ast}\left(  \mathbf{r},t\right)
,\label{E}\\
\mathbf{E}\left(  \mathbf{r},t\right)   &  =\int d^{3}k\,\mathbf{E}\left(
\mathbf{k}\right)  e^{i\left(  \mathbf{k}\cdot\mathbf{r}-ckt\right)  },
\end{align}
interacting with a closed two--level atom or molecule, whose excited and
fundamental states are $\left\vert 2\right\rangle $ and $\left\vert
1\right\rangle $, respectively. The two--level atom has a transition frequency
$\omega_{21}$ and electric dipole matrix elements $\left\langle 2\right\vert
\mathbf{\hat{\mu}}\left\vert 1\right\rangle =\left\langle 1\right\vert
\mathbf{\hat{\mu}}\left\vert 2\right\rangle \equiv\mu\mathbf{z}$, which have
been taken to be real vectors, alligned parallel to the cartesian $z-$axis,
without loss of generality. In the Dirac picture, and after performing the
rotating--wave approximation, the semiclassical optical Bloch equations for
an\textit{ individual atom} located at point $\mathbf{r}$ can be written as
\cite{Shore,Milonni,Loudon}
\begin{align}
\dot{\rho}_{22}\left(  \mathbf{r},t\right)   &  =-A\rho_{22}-i\left[
\Omega^{\ast}\left(  \mathbf{r},t\right)  \rho_{21}-\Omega\left(
\mathbf{r},t\right)  \rho_{12}\right]  ,\label{b1}\\
\dot{\rho}_{11}\left(  \mathbf{r},t\right)   &  =+A\rho_{22}+i\left[
\Omega^{\ast}\left(  \mathbf{r},t\right)  \rho_{21}-\Omega\left(
\mathbf{r},t\right)  \rho_{12}\right]  ,\label{b2}\\
\dot{\rho}_{21}\left(  \mathbf{r},t\right)   &  =-\frac{1}{2}A\rho
_{21}-i\Omega\left(  \mathbf{r},t\right)  \left(  \rho_{22}-\rho_{11}\right)
, \label{b3}%
\end{align}
where $\rho_{22}\left(  \rho_{11}\right)  $ is the population of the upper
(lower) atomic level, $\rho_{12}$ is the slowly varying atomic coherence and
\begin{equation}
\Omega\left(  \mathbf{r},t\right)  =\frac{\mu}{\hbar}\mathbf{z}\cdot
\mathbf{E}\left(  \mathbf{r},t\right)  e^{i\omega_{21}t}, \label{Rabi}%
\end{equation}
is half the complex Rabi frequency of the light field. The effect of
spontaneous emission needs to be described phenomenologicaly (through the
damping/pumping terms proportional to the $A$ coefficient) since, as it is
well known, the standard semiclassical theory cannot describe spontaneous
emission \cite{note}.

The optical Bloch equations (\ref{b1})--(\ref{b3}) admit the simpler form
\begin{align}
\dot{n}  &  =-A\left(  n+1\right)  -2i\left(  \Omega^{\ast}\rho_{21}%
-\Omega\rho_{12}\right)  ,\label{B1}\\
\dot{\rho}_{21}  &  =-\frac{1}{2}A\rho_{21}-i\Omega n, \label{B2}%
\end{align}
where $n\left(  \mathbf{r},t\right)  \equiv\rho_{22}\left(  \mathbf{r}%
,t\right)  -\rho_{11}\left(  \mathbf{r},t\right)  $ is the population
inversion of the atom.

Now Eq.(\ref{B2}) can be integrated formally yielding
\begin{equation}
\rho_{21}\left(  \mathbf{r},t\right)  =-i\int_{0}^{t}d\tau\ \Omega\left(
\mathbf{r},\tau\right)  n\left(  \mathbf{r},\tau\right)  e^{\frac{1}%
{2}A\left(  \tau-t\right)  },
\end{equation}
where we have taken $\rho_{21}\left(  \mathbf{r},0\right)  =0$ for the sake of
simplicity. Substituting this into Eq.(\ref{B1}) we get
\begin{align}
\dot{n}  &  =-A\left(  n+1\right) \label{npunt}\\
&  -4\operatorname{Re}\int_{0}^{t}d\tau\ \Omega^{\ast}\left(  \mathbf{r}%
,t\right)  \Omega\left(  \mathbf{r},\tau\right)  n\left(  \mathbf{r}%
,\tau\right)  e^{\frac{1}{2}A\left(  \tau-t\right)  }.\nonumber
\end{align}

\subsection{Ensemble averaging}

Let us emphasize that Eq.(\ref{npunt}) is the equation of evolution for the
population inversion \textit{for a single} atom. As we are interested in the
average evolution of the whole system, the ensemble of atoms, which is the
quantity whose evolution ERE describe, we next write down the evolution
equation for the ensemble averaged inversion
\begin{equation}
\bar{n}\left(  t\right)  \equiv\left\langle n\left(  \mathbf{r}_{a},t\right)
\right\rangle ,
\end{equation}
where the index $a$ labels each atom and
\begin{equation}
\left\langle f\left(  \mathbf{r}_{a},t\right)  \right\rangle =\frac{1}{N}%
\sum_{a=1}^{N}f\left(  \mathbf{r}_{a},t\right)  , \label{av}%
\end{equation}
for any arbitrary function $f$ $\left(  \mathbf{r}_{a},t\right)  $ ($N$ the
total number ot atoms). The equation of evolution of the ensemble--averaged
inversion $\bar{n}\left(  t\right)  $ is obtained from Eq. (\ref{npunt}) and
reads
\begin{align}
\frac{d}{dt}\bar{n}  &  =-A\left(  \bar{n}+1\right)  -4\operatorname{Re}%
\int_{0}^{t}d\tau C_{n}\left(  t,\tau\right)  e^{\frac{1}{2}A\left(
\tau-t\right)  },\label{npunt2}\\
C_{n}\left(  t,\tau\right)   &  =\left\langle \Omega^{\ast}\left(
\mathbf{r}_{a},t\right)  \Omega\left(  \mathbf{r}_{a},\tau\right)  n\left(
\mathbf{r}_{a},\tau\right)  \right\rangle . \label{corrn}%
\end{align}
In the next section we derive ERE for the particular case of a weak light
field, and leave the general case for the following section.

\section{Derivation of the Einstein's rate equations I: Weak field}

We start with Eq.(\ref{npunt}) and follow the same steps as in deriving
Eq.(\ref{Einstein weak}) in subsection \ref{weak}: Assume a weak field
$\Omega\left(  \mathbf{r},\tau\right)  =\varepsilon\Omega_{1}\left(
\mathbf{r},\tau\right)  $, expand the inversion in powers of $\varepsilon$,
Eq. (\ref{desarrollo}), and find the equation of evolution at order
$\varepsilon$. One easily gets%
\begin{equation}
\dot{n}_{1}=-A\left(  n_{1}+1\right)  +4\operatorname{Re}\int_{0}^{t}%
d\tau\Omega^{\ast}\left(  \mathbf{r},t\right)  \Omega\left(  \mathbf{r}%
,\tau\right)  e^{\frac{1}{2}A\left(  \tau-t\right)  }.
\end{equation}
Now, performing the ensemble averaging%
\begin{equation}
\frac{d}{dt}\bar{n}_{1}=-A\left(  \bar{n}_{1}+1\right)  +4\operatorname{Re}%
\int_{0}^{t}d\tau C\left(  t,\tau\right)  e^{\frac{1}{2}A\left(
\tau-t\right)  }, \label{n1punt}%
\end{equation}
where $C\left(  t,\tau\right)  $ is the correlation function
\begin{equation}
C\left(  t,\tau\right)  =\left\langle \Omega^{\ast}\left(  \mathbf{r}%
_{a},t\right)  \Omega\left(  \mathbf{r}_{a},\tau\right)  \right\rangle .
\label{c}%
\end{equation}

\subsection{Evaluation of the correlation $C\left(  t,\tau\right)  $}

In this subsection we calculate the correlation function $C\left(
t,\tau\right)  $ for the special case of an isotropic broadband light--field.
From Eq. (\ref{c}) and definition (\ref{av}) we have%

\begin{equation}
C\left(  t,\tau\right)  =\frac{1}{N}\sum_{a=1}^{N}\Omega^{\ast}\left(
\mathbf{r}_{a},t\right)  \Omega\left(  \mathbf{r}_{a},\tau\right)  ,
\label{c1}%
\end{equation}
which, under the assumptions of very large $N$ and homogeneous spatial
distribution of atoms, can be computed as
\begin{equation}
C\left(  t,\tau\right)  =\frac{1}{V}\int d^{3}r\ \Omega^{\ast}\left(
\mathbf{r},t\right)  \Omega\left(  \mathbf{r},\tau\right)  . \label{c2}%
\end{equation}
with $V$ the volume occupied by the sample. Making use of Eq. (\ref{Rabi}),
Eq. (\ref{c2}) becomes
\begin{equation}
C\left(  t,\tau\right)  =\frac{\mu^{2}}{V\hbar^{2}}\int d^{3}r\ E_{z}^{\ast
}\left(  \mathbf{r},t\right)  E_{z}\left(  \mathbf{r},\tau\right)
e^{-i\omega_{21}\left(  t-\tau\right)  }, \label{c3}%
\end{equation}
with $E_{z}\left(  \mathbf{r},t\right)  =\mathbf{z}\cdot\mathbf{E}\left(
\mathbf{r},t\right)  $. By using Eq. (\ref{E}), and assuming a large enough
gas volume as for being possible to apply $\int d^{3}r\ e^{i\left(
\mathbf{k}^{\prime}-\mathbf{k}\right)  \cdot\mathbf{r}}=\left(  2\pi\right)
^{3}\delta\left(  \mathbf{k}^{\prime}-\mathbf{k}\right)  $, which is strictly
valid only for an infinite volume, Eq. (\ref{c3}) becomes
\begin{equation}
C\left(  t,\tau\right)  =\frac{\left(  2\pi\right)  ^{3}\mu^{2}}{V\hbar^{2}%
}e^{-i\omega_{21}\left(  t-\tau\right)  }\int d^{3}k\ \left\vert E_{z}\left(
\mathbf{k}\right)  \right\vert ^{2}e^{ick\left(  t-\tau\right)  }, \label{c4}%
\end{equation}
with $E_{z}\left(  \mathbf{k}\right)  =\mathbf{z}\cdot\mathbf{E}\left(
\mathbf{k}\right)  $

We now evaluate the integral in Eq.(\ref{c4}) by expressing the electric
vectors as
\begin{equation}
\mathbf{E}\left(  \mathbf{k}\right)  =\mathbf{e}_{||}\left(  \mathbf{k}%
\right)  E_{||}\left(  \mathbf{k}\right)  +\mathbf{e}_{\perp}\left(
\mathbf{k}\right)  E_{\perp}\left(  \mathbf{k}\right)  , \label{epol}%
\end{equation}
where the polarization unit vectors $\mathbf{e}_{||}\left(  \mathbf{k}\right)
$ and $\mathbf{e}_{\perp}\left(  \mathbf{k}\right)  $ are perpendicular to
$\mathbf{k}$, and are chosen to be parallel and orthogonal, respectively, to
the plane defined by $\mathbf{k}$ and $\mathbf{z}$. With this choice,
\begin{equation}
\mathbf{z}\cdot\mathbf{e}_{||}\left(  \mathbf{k}\right)  =\sin\theta
,\quad\mathbf{z}\cdot\mathbf{e}_{\perp}\left(  \mathbf{k}\right)  =0,
\label{scalar}%
\end{equation}
with $\theta$ is the angle defined by $\mathbf{k}$ and $\mathbf{z}$
\cite{note2}.

Substituting Eqs. (\ref{epol}) and (\ref{scalar}) into Eq. (\ref{c4}) and
expressing the integral in spherical coordinates, we obtain
\begin{align}
C\left(  t,\tau\right)   &  =\frac{\left(  2\pi\right)  ^{3}\mu^{2}}%
{V\hbar^{2}}e^{-i\omega_{21}\left(  t-\tau\right)  }\int_{0}^{2\pi}%
d\varphi\int_{0}^{\pi}d\theta\sin^{3}\theta\label{c5}\\
&  \times\int_{0}^{\infty}dk\ k^{2}\left\vert E_{||}\left(  \mathbf{k}\right)
\right\vert ^{2}e^{ick\left(  t-\tau\right)  }.\nonumber
\end{align}
We finally assume that $\left\vert E\left(  \mathbf{k}\right)  \right\vert
^{2}$ is dependent on $k$ but not on the orientation of $\mathbf{k}$, i.e. we
assume that the light energy at a given $k$ is distributed isotropically in
all directions. In particular, we write
\begin{equation}
\left\vert E_{||}\left(  \mathbf{k}\right)  \right\vert ^{2}=\left\vert
E_{||}\left(  \omega\right)  \right\vert ^{2},\quad\omega=ck. \label{eomega}%
\end{equation}
Under this assumption, the angular integrations can be done and changing the
integration varible from $k$ to $\omega$, Eq. (\ref{c5}) becomes
\begin{align}
C\left(  t,\tau\right)   &  =\frac{\left(  2\pi\right)  ^{4}4\mu^{2}}%
{3V\hbar^{2}c^{3}}e^{-i\omega_{21}\left(  t-\tau\right)  }\label{c6}\\
&  \times\int_{0}^{\infty}d\omega\ \omega^{2}\left\vert E_{||}\left(
\omega\right)  \right\vert ^{2}e^{i\omega\left(  t-\tau\right)  }.\nonumber
\end{align}
Finally, taking into account that the spectral density of energy per unit
volume of the light field can be written as (see the Appendix)
\begin{equation}
W\left(  \omega\right)  =\frac{8\varepsilon_{0}}{Vc^{3}}\left(  2\pi\right)
^{4}\omega^{2}\,\left\vert E_{||}\left(  \omega\right)  \right\vert ^{2},
\label{W}%
\end{equation}
we can express the correlation function $C\left(  t,\tau\right)  $, Eq.
(\ref{c6}), in terms of $W\left(  \omega\right)  $ as
\begin{equation}
C\left(  t,\tau\right)  =\frac{\mu^{2}}{6\hbar^{2}\varepsilon_{0}}%
e^{-i\omega_{21}\left(  t-\tau\right)  }\int_{0}^{\infty}d\omega\ W\left(
\omega\right)  e^{i\omega\left(  t-\tau\right)  }. \label{c6bis}%
\end{equation}
Introducing $\beta=\omega-\omega_{21}$, we get
\begin{equation}
C\left(  t,\tau\right)  =\frac{\mu^{2}}{6\hbar^{2}\varepsilon_{0}}%
\int_{-\infty}^{+\infty}d\beta\ W\left(  \omega_{21}+\beta\right)
e^{i\beta\left(  t-\tau\right)  }. \label{c7}%
\end{equation}
where the lower integration limit has been extended from $-\omega_{21}$ to
$-\infty$, which is a very good approximation for optical frequencies. Hence
it turns out that $C\left(  t,\tau\right)  $ is proportional to the Fourier
transform of the spectral density of energy of the light field.

\subsection{\label{re}Rate Equations}

Making use of Eq.(\ref{c7}), Eq.(\ref{n1punt}) reads
\begin{align}
\frac{d}{dt}\bar{n}_{1}\left(  t\right)   &  =-A\left(  \bar{n}_{1}+1\right)
+\frac{2\mu^{2}}{3\hbar^{2}\varepsilon_{0}}\label{dn1dt}\\
&  \times\int_{-\infty}^{+\infty}d\beta\ W\left(  \omega_{21}+\beta\right)
K\left(  \beta\right)  ,\nonumber\\
K\left(  \beta\right)   &  =\operatorname{Re}\int_{0}^{t}d\tau e^{-\left(
\frac{1}{2}A-i\beta\right)  \left(  t-\tau\right)  }.
\end{align}
Performing the integral $K\left(  \beta\right)  $ one gets,
\begin{equation}
K\left(  \beta\right)  =\operatorname{Re}\frac{1-e^{-\left(  \frac{1}%
{2}A-i\beta\right)  t}}{\frac{1}{2}A-i\beta}\overset{At\gg1}{\rightarrow}%
\frac{\frac{1}{2}A}{\left(  \frac{1}{2}A\right)  ^{2}+\beta^{2}}, \label{K}%
\end{equation}%
\[
\operatorname{Re}\int_{-\infty}^{+\infty}d\beta\ W\left(  \omega_{21}%
+\beta\right)  K\approx\pi W\left(  \omega_{21}\right)
\]
which is a Lorentzian of width $A$ centered ar $\beta=0$. Going back to Eq.
(\ref{dn1dt}), if $W\left(  \omega_{21}+\beta\right)  $ is a smooth function,
(specifically, if $A\left[  \partial W\left(  \omega_{21}+\beta\right)
/\partial\beta\right]  _{\beta=0}\ll1$) we can take $W\left(  \omega
_{21}+\beta\right)  =W\left(  \omega_{21}\right)  $ and the final result reads
($\int_{-\infty}^{+\infty}d\beta\ K\left(  \beta\right)  =\pi$):
\begin{equation}
\frac{d}{dt}\bar{n}_{1}\left(  t\right)  =-A\left(  \bar{n}_{1}+1\right)
+\frac{2\pi\mu^{2}}{3\hbar^{2}\varepsilon_{0}}W\left(  \omega_{21}\right)  ,
\end{equation}
which compared with Eq.(\ref{Einstein weak}) provides the following value for
the Einstein $B$ coefficient:%
\begin{equation}
B=\frac{\pi\mu^{2}}{3\hbar^{2}\varepsilon_{0}}, \label{Bcoef}%
\end{equation}
which is the correct one for isotropic radiation \cite{Loudon}.

\section{Derivation of the Einstein's rate equations II: Arbitrary field
strength}

Now we remove the weak light field assumption and return to Eq.(\ref{npunt2}).
Now it is the correlation $C_{n}$, Eq.(\ref{corrn}), that has to be evaluated.

We are considering broadband incoherent light fields. This implies that the
quantities appearing in $C_{n}$ oscillate randomly in time: On the one hand
the phases of the Fourier components components of the field, $\mathbf{E}%
\left(  \mathbf{k}\right)  $, are random as the field is incoherent, and on
the other hand, the values taken by the population inversion of a single atom
$n\left(  \mathbf{r},\tau\right)  $ are not correlated with those of the
fields at this very instant, because the value of the inversion depends on
previous times. Thus, it seems reasonable to assume that the random variations
of the inversion of each individual atom, $n\left(  \mathbf{r}_{a}%
,\tau\right)  $, are decorrelated from the random variations of the mutual
intensity $\Omega^{\ast}\left(  \mathbf{r}_{a},t\right)  \Omega\left(
\mathbf{r}_{a},\tau\right)  $. This assumption constitutes our main
approximation. Thus, we assume%

\begin{equation}
C_{n}\left(  t,\tau\right)  \approx\left\langle \Omega^{\ast}\left(
\mathbf{r}_{a},t\right)  \Omega\left(  \mathbf{r}_{a},\tau\right)
\right\rangle \left\langle n\left(  \mathbf{r}_{a},\tau\right)  \right\rangle
=C\left(  t,\tau\right)  \bar{n}\left(  \tau\right)  , \label{decorr}%
\end{equation}
with $C\left(  t,\tau\right)  $ given by Eq.(\ref{c7}).

Once Eq.(\ref{decorr}) is assumed, the derivation of ERE is easy: substituting
Eqs.(\ref{decorr}) and (\ref{c7}) into Eq. (\ref{npunt2}) one gets
\begin{align}
\frac{d}{dt}\bar{n}\left(  t\right)   &  =-A\left[  \bar{n}\left(  t\right)
+1\right]  -\frac{2\mu^{2}}{3\hbar^{2}\varepsilon_{0}}\operatorname{Re}%
\int_{0}^{t}d\tau\bar{n}\left(  \tau\right)  e^{-\frac{A}{2}\left(
t-\tau\right)  }\\
&  \int_{-\infty}^{+\infty}d\beta\ W\left(  \omega_{21}+\beta\right)
e^{i\beta\left(  t-\tau\right)  },\nonumber
\end{align}
which after making the same assumptions as in Sec. \ref{re} becomes
\begin{equation}
\frac{d}{dt}\bar{n}\left(  t\right)  =-A\left[  \bar{n}\left(  t\right)
+1\right]  -\frac{2\pi\mu^{2}}{3\hbar^{2}\varepsilon_{0}}W\left(  \omega
_{21}\right)  \bar{n}\left(  t\right)  .
\end{equation}
This is the same Eq. (\ref{Einstein}) after identifying the $B$ coefficient,
Eq.(\ref{Bcoef}).

We see that it is not necessary to assume a weak field for deriving the $B$
coefficient with the semiclassical theory. In spite of that, for arbitrary
field intensity it is the statistical independence of the flutuations of the
inversion and the mutual intensity that must be invoked.

\section{Estimation of the domain of applicability of the Einstein's rate
equations}

The derivation presented above is a bit crude in the sense that the
approximation carried out in Sect. \ref{re} (that $W\left(  \omega\right)  $
was a smooth enough function of $\omega$) does not inform on the domain of
validity of the resulting ERE. In other words, we have assumed an infinitely
broad spectrum for the light field and the natural question is: How broad must
the spectrum be for ERE be valid? In this section we give such an estimation
by considering a finite spectrum linewidth. Especifically, we assume that
$W\left(  \omega\right)  $ has a Lorentzian form, which represents many actual
light sources but can be regarded also as a representation of any bell-shaped
distribution. Then we take
\begin{equation}
W\left(  \omega\right)  =W\left(  \omega_{0}\right)  \frac{\gamma^{2}}%
{\gamma^{2}+\left(  \omega-\omega_{0}\right)  ^{2}}, \label{lorentzian}%
\end{equation}
where $\omega_{0}$ is the center of the spectrum, which has a width $\gamma$.
We note that
\begin{equation}
W\left(  \omega_{21}\right)  =W\left(  \omega_{0}\right)  \frac{\gamma^{2}%
}{\gamma^{2}+\delta^{2}}, \label{W21}%
\end{equation}
where $\delta=\omega_{21}-\omega_{0}$ is the detuning between the transition
frequency and the center of the light field spectrum. With this lorentzian
form, Eq. (\ref{c6bis}) can be evaluated to yield
\begin{equation}
C\left(  t,\tau\right)  =\frac{\pi\mu^{2}}{6\hbar^{2}\varepsilon_{0}}\gamma
W\left(  \omega_{0}\right)  e^{-i\delta\left(  t-\tau\right)  }e^{-\gamma
\left(  t-\tau\right)  }. \label{c8}%
\end{equation}
We now go back to Eq. (\ref{npunt2}), with approximation (\ref{decorr}), and
obtain
\begin{align}
\frac{d}{dt}\bar{n}\left(  t\right)   &  =-A\left[  \bar{n}\left(  t\right)
+1\right]  -2\gamma BW\left(  \omega_{0}\right) \label{npunt4}\\
&  \times\operatorname{Re}\int_{0}^{t}d\tau~\bar{n}\left(  \tau\right)
e^{-\left(  \frac{1}{2}A+\gamma+i\delta\right)  \left(  t-\tau\right)
},\nonumber
\end{align}
where we have made use of Eq. (\ref{Bcoef}). We finally evaluate the integral
by parts repeatedly, yielding
\begin{align}
\frac{d}{dt}\bar{n}\left(  t\right)   &  =-A\left[  \bar{n}\left(  t\right)
+1\right]  -2\gamma BW\left(  \omega_{0}\right) \label{npunt5}\\
&  \operatorname{Re}\sum_{p=0}^{\infty}\frac{\left(  -1\right)  ^{p}}{\left(
\frac{1}{2}A+\gamma+i\delta\right)  ^{p+1}}\nonumber\\
&  \times\left[  \bar{n}^{\left(  p\right)  }\left(  t\right)  -\bar
{n}^{\left(  p\right)  }\left(  0\right)  e^{-\left(  \frac{1}{2}%
A+\gamma+i\delta\right)  t}\right]  ,\nonumber
\end{align}
where the superscript $\left(  p\right)  $ denotes the $p-$th derivative with
respect to $t$. If we now assume that $\gamma\gg A$ (the spectrum is broad
referred to the natural linewidth of the atomic transition), the exponentials
in Eq. (\ref{npunt5}) can be ignored as they are damped out in a time
$\sim\gamma^{-1}$ much smaller that the characteristic time $A^{-1}$ of the
inversion dynamics. Then
\begin{align}
\frac{d}{dt}\bar{n}\left(  t\right)   &  =-A\left[  \bar{n}\left(  t\right)
+1\right]  -2\gamma BW\left(  \omega_{0}\right)  \sum_{p=0}^{\infty}%
S_{p}\left(  t\right)  ,\label{npunt6}\\
S_{p}\left(  t\right)   &  =\bar{n}^{\left(  p\right)  }\left(  t\right)
\operatorname{Re}\frac{\left(  -1\right)  ^{p}}{\left(  \gamma+i\delta\right)
^{p+1}}.
\end{align}

Clearly, in order that Eq. (\ref{npunt6}) be equivalent to Einstein equation,
Eq. (\ref{Einstein}), the terms with $p\geq1$ should be negligible. Let us
then evaluate the order of magnitude of the $p-$th order, which we denote by
$\mathcal{O}\left(  S_{p}\right)  $:
\begin{align}
\mathcal{O}\left(  S_{p}\right)   &  =\mathcal{O}\left(  \bar{n}^{\left(
p\right)  }\right)  \operatorname{Re}\frac{1}{\left(  \gamma+i\delta\right)
^{p+1}}\nonumber\\
&  \leq\mathcal{O}\left(  \bar{n}^{\left(  p\right)  }\right)  \left\vert
\gamma+i\delta\right\vert ^{-\left(  p+1\right)  }=\mathcal{O}\left(  \bar
{n}^{\left(  p\right)  }\right)  \left(  \gamma^{2}+\delta^{2}\right)
^{-\frac{p+1}{2}}. \label{Osp}%
\end{align}
We thus need to evaluate $\mathcal{O}\left(  \bar{n}^{\left(  p\right)
}\right)  $. We start with $p=0$, i.e., $\bar{n}^{\left(  0\right)  }\left(
t\right)  =\bar{n}\left(  t\right)  $ whose order of magnitude is $1$. Thus
$\mathcal{O}\left(  \bar{n}^{\left(  0\right)  }\right)  =1$. We proceed with
$p=1$, and assume that it can be computed by neglecting the $S_{p\geq1}$'s in
Eq. (\ref{npunt5}). We then have
\begin{equation}
\bar{n}^{\left(  1\right)  }\left(  t\right)  =-A\left[  \bar{n}\left(
t\right)  +1\right]  -2BW\left(  \omega_{21}\right)  \bar{n}\left(  t\right)
, \label{O1}%
\end{equation}
where Eq. (\ref{W21}) has been used. An upper bound to the order of magnitude
of $\bar{n}^{\left(  1\right)  }$ can be estimated by using $\mathcal{O}%
\left(  \bar{n}^{\left(  0\right)  }\right)  \equiv\mathcal{O}\left(  \bar
{n}\right)  =1$, hence $\mathcal{O}\left(  \bar{n}^{\left(  1\right)
}\right)  =2A+2BW\left(  \omega_{21}\right)  $. Similarly we obtain $\bar
{n}^{\left(  2\right)  }\left(  t\right)  \simeq-\left[  A+2BW\left(
\omega_{21}\right)  \right]  \bar{n}^{\left(  1\right)  }\left(  t\right)  $,
and making use of Eq. (\ref{O1}), $\mathcal{O}\left(  \bar{n}^{\left(
2\right)  }\right)  =\left[  2A+2BW\left(  \omega_{21}\right)  \right]  ^{2}$,
where a factor $2$ multiplying $A$ has beed added, which amounts to
overestimate $\mathcal{O}\left(  \bar{n}^{\left(  2\right)  }\right)  $. In
general one gets $\mathcal{O}\left(  \bar{n}^{\left(  p\right)  }\right)
=\left[  2A+2BW\left(  \omega_{21}\right)  \right]  ^{p}$. Substituting this
expression into Eq. (\ref{Osp}) we obtain
\begin{equation}
\mathcal{O}\left(  S_{p}\right)  =\gamma\left[  2A+2BW\left(  \omega
_{21}\right)  \right]  ^{p}\left(  \gamma^{2}+\delta^{2}\right)  ^{-\frac
{p+1}{2}}.
\end{equation}
Now we now impose $\mathcal{O}\left(  S_{p+1}\right)  \ll\mathcal{O}\left(
S_{p}\right)  $, what directly leads to an upper bound to the spectral
density,
\begin{equation}
\frac{2BW\left(  \omega_{21}\right)  }{\sqrt{\gamma^{2}+\delta^{2}}}\ll1,
\label{bound}%
\end{equation}
where we have recalled that $A\ll\gamma$.

In resume, under condition (\ref{bound}), all $S_{p\geq1}$ can be neglected
and Eq. (\ref{npunt6}) reduces to the Einstein's rate equation Eq.
(\ref{Einstein}). Notice that condition (\ref{bound}) imposes a constrain on
the field spectrum for ERE being applicable: If the energy density $W\left(
\omega_{21}\right)  $ is very large, the spectrum linewidth $\gamma$ (or the
detuning $\delta$) must be consequently large.

\section{Conclusions}

We have provided a derivation of the Einstein's rate equations from the
semiclassical Bloch equation for an esemble of two--level atoms or molecules.
The derivation has been done first for a weak field and we have obtained the
right expression for the $B$ coefficient. In our derivation the necessary
ensemble averaging is done explicitly. Then, we have generalized the
derivation to the case or arbitrary field strength by assuming a statistical
independence of the flutuations of the atomic inversion and the mutual
intensity, which is a reasonable assumption for an incoherent light field.
Finally we have estimated which condition the spectrum of such an incoherent
field must verify for the ERE be valid, Eq. (\ref{bound}).

Let us finally note that in all the above derivations the condition $At\gg1$
has been invoked. Thus the ERE describe the "long term" behavior of the system
and their solutions are only meaningful in this limit.

\section{Appendix}

Let us now see how Eq. (\ref{c6}) relates to the spectral density of energy
per unit volume of the light field. The energy per unit volume $\eta$ can be
computed as%
\begin{equation}
\eta=\frac{\varepsilon_{0}}{2V}\int d^{3}r\left[  \mathcal{\vec{E}}^{2}\left(
\mathbf{r},t\right)  +c^{2}\mathcal{\vec{B}}^{2}\left(  \mathbf{r},t\right)
\right]  .
\end{equation}
By making use of Eq. (\ref{E}), writing $\mathcal{\vec{B}}\left(
\mathbf{r},t\right)  =\mathbf{B}\left(  \mathbf{r},t\right)  +\mathbf{B}%
^{\ast}\left(  \mathbf{r},t\right)  $ with%
\begin{equation}
\mathbf{B}\left(  \mathbf{r},t\right)  =\int d^{3}k\,\mathbf{B}\left(
\mathbf{k}\right)  e^{i\left(  \mathbf{k}\cdot\mathbf{r}-ckt\right)  },
\end{equation}
and making use of the well known property $\left(  \mathbf{a}\times
\mathbf{b}\right)  \cdot\left(  \mathbf{c}\times\mathbf{d}\right)  =\left(
\mathbf{a}\cdot\mathbf{c}\right)  \left(  \mathbf{b\cdot d}\right)  -\left(
\mathbf{a\cdot d}\right)  \left(  \mathbf{b\cdot c}\right)  $, it can be
readily shown that%
\begin{equation}
\eta=2\frac{\varepsilon_{0}}{V}\left(  2\pi\right)  ^{3}\int d^{3}k\,\left[
\left\vert \mathbf{E}\left(  \mathbf{k}\right)  \right\vert ^{2}%
+c^{2}\left\vert \mathbf{B}\left(  \mathbf{k}\right)  \right\vert ^{2}\right]
. \label{etaaux}%
\end{equation}
Taking into account that%
\begin{equation}
\mathbf{B}\left(  \mathbf{k}\right)  =\frac{\mathbf{k\times E}\left(
\mathbf{k}\right)  }{c\left\vert \mathbf{k}\right\vert },
\end{equation}
Eq. (\ref{etaaux}) simplifies to%
\begin{equation}
\eta=2\frac{\varepsilon_{0}}{V}\left(  2\pi\right)  ^{3}\int d^{3}k\left\vert
\mathbf{E}\left(  \mathbf{k}\right)  \right\vert ^{2}.
\end{equation}
Now, making use of Eq. (\ref{epol}), expressing this integral in spherical
coordinates, and changing the integration variable from $k$ to $\omega=ck$,
one gets
\begin{equation}
\eta=8\frac{\varepsilon_{0}}{Vc^{3}}\left(  2\pi\right)  ^{4}\int_{0}^{\infty
}d\omega\omega^{2}\left\vert E_{||}\left(  \omega\right)  \right\vert ^{2},
\label{eta3}%
\end{equation}
where the angular integration has been performed and again it has been used
the isotropy condition $\left\vert E_{\perp}\left(  \omega\right)  \right\vert
^{2}=\left\vert E_{||}\left(  \omega\right)  \right\vert ^{2}$. Finally, from
Eq. (\ref{eta3}) we identify the spectral density of energy per unit volume as%
\begin{equation}
W\left(  \omega\right)  =8\frac{\varepsilon_{0}}{Vc^{3}}\left(  2\pi\right)
^{4}\omega^{2}\left\vert E_{||}\left(  \omega\right)  \right\vert ^{2}.
\label{Wapp}%
\end{equation}


\begin{thebibliography}{99}
\bibitem {note0}Let us stress that derivations of rate equations through
adibatic ellimination of the medium polarization (valid in the limit of large
decay for the atomic coherence, as compared to that of the population
inversion, see e.g. \cite{Milonni}) are not true derivations of Einstein's
rate equations. For example, in this way one does not obtain the right
expresion for the Einstein's coefficients. For deriving ERE, one must take
into account the broadband nature of the light field and has not to imposse
any a priori condition on the atomic decay rates.
\bibitem {note2}Note that the single case in which the referred plane is
undefined is when $\mathbf{k}$ and $\mathbf{z}$ are parallel, i.e. $\theta=0$,
but in this case both polarization unit vectors are orthogonal to $\mathbf{z}%
$, in agrement with the result $\mathbf{z}\cdot\mathbf{e}_{||}\left(
\mathbf{k}\right)  =\sin\theta$ $=0$ and $\mathbf{z}\cdot\mathbf{e}_{\perp
}\left(  \mathbf{k}\right)  =0$.
\end{thebibliography}
\end{document}